\documentclass[seceq]{ptptex}
\usepackage{wrapft}
\usepackage{graphicx}
\def\be{\begin{equation}}
\def\ee{\end{equation}}
\def\bea{\begin{eqnarray}}
\def\eea{\end{eqnarray}}

\pubinfo{Vol.~111, No.~4, April 2004}

\title{
Moduli Stabilization in String Gas Cosmology
}
\subtitle{\LaTeXe\ Version}    

\author{
Robert H. \textsc{Brandenberger}\footnote{E-mail: rhb@hep.physics.mcgill.ca} 
}

\inst{
Physics Department, McGill University, Montreal, QC, H3T 2A8, Canada}

\recdate{
October 1, 2005}

\abst{
String gas cosmology is an approach towards studying the effects of
superstring theory on early universe cosmology which is based on new
symmetries and new degrees of freedom of string theory. Within this
context, it appears possible to stabilize the moduli which describe
the size and shape of the extra spatial dimensions without the need
of introducing many extra tools such as warping and fluxes. In this
lecture, the recent progress towards moduli stabilization in string
gas cosmology is reviewed, and outstanding problems for the scenario
are discussed.
}

\begin{document}

\maketitle

\section{Introduction}

In this Introduction, the conceptual problems
of scalar field-driven inflationary cosmology (our
current paradigm of early universe cosmology) will be
reviewed. These are some of the reasons which
motivate the search for a new theory of the very early universe.
It will then be explained why string theory offers the promise
of successfully addressing some of these problems.
However, any attempt to combine string theory with cosmology
leads to new cosmological problems, most importantly the
questions of why only three of the nine spatial dimensions of
critical superstring theory are large, and why the moduli fields
describing the volume and shape of the extra dimensions are
stabilized.  

In the context of the current view of cosmology in which
the temperature increases without bound as we venture into
the past, it is clear that it is the ultimate theory which
describes physics at the highest energies and on the smallest
scales which will determine the evolution of the universe at the
earliest times. At the moment, string theory is the best
candidate of providing the required theory of matter at the
highest energy scales. Thus, in the following we shall work under
the hypothesis that string theory is indeed the correct theory
of matter in the very early universe.

However, there are also reasons coming from cosmology alone which
might lead us to turn to string theory in the search for the
theory of the very early universe: Most implementations of
the inflationary universe scenario \cite{Guth,Sato} 
(see also \cite{Starob1,Brout} for earlier ideas),
the current paradigm of early universe cosmology, are based
on the existence of a slowly rolling scalar field, the
{\it inflaton}, whose energy-momentum
tensor is dominated by the contribution of the field potential energy 
which drives a period of accelerated expansion. 
In spite of the impressive phenomenological
successes of this paradigm in predicting the spectrum of density
perturbations and the angular power spectrum of cosmic microwave
background (CMB) anisotropies, the scalar field-driven inflationary
universe scenario suffers from some important conceptual problems
\cite{RHBrev3,Review1}.
Addressing these problems is another of the goals of superstring
cosmology.

The first of the conceptual problems of scalar field-driven inflation
is the {\it amplitude problem}, namely the
fact that in the simplest realizations of the model, a hierarchy
of scales needs to be present in order to be able to obtain the
observed small amplitude of the primordial anisotropies (see
e.g. \cite{Adams} for a fairly general analysis of this problem).

A more serious problem is the {\it trans-Planckian problem} \cite{RHBrev3}.
As can be seen from the space-time diagram of Figure 1, 
provided that the period of inflation lasted sufficiently
long (for GUT scale inflation the number is about 70 e-foldings),
then all scales inside the Hubble radius today started out with a
physical wavelength smaller than the Planck scale at the beginning of
inflation. Now, the current theory of cosmological perturbations
(the theory used to calculate the spectra of density fluctuations
and microwave anisotropies) is based
on Einstein's theory of General Relativity coupled to a simple
semi-classical description of matter. It is clear that these
building blocks of the theory are inapplicable on scales comparable
and smaller than the Planck scale. Thus, the key
successful prediction of inflation (the theory of the origin of
fluctuations) is based on an incomplete analysis: we know that 
new physics must enter
into a correct computation of the spectrum of cosmological perturbations.
The key question is whether the predictions obtained using
the current theory are sensitive to the specifics of the unknown
theory which takes over on small scales. Toy model calculations
using modified dispersion relations \cite{Jerome1,Niemeyer} have
shown that the predictions are in fact sensitive to Planck-scale
physics, thus opening up the exciting possibility to test Planck-scale
and string-scale physics in current observations (see \cite{Jerome2}
for a review with references to other approaches towards exploring
this ``trans-Planckian window of opportunity'').
\begin{figure}
       \centerline{\includegraphics[width=6 cm,height=6 cm]
                                   {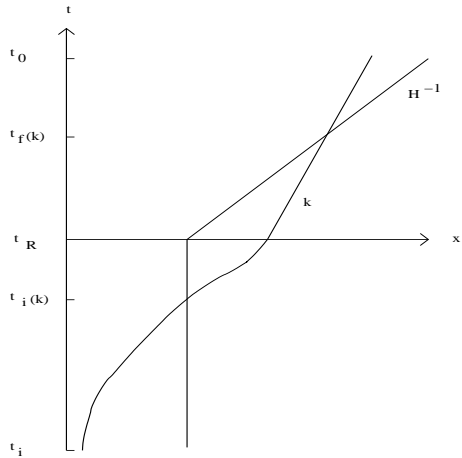}}
   \caption{Space-time diagram (sketch) showing the evolution
of scales in inflationary cosmology. The vertical axis is
time, and the period of inflation lasts between $t_i$ and
$t_R$, and is followed by the radiation-dominated phase
of standard big bang cosmology. During exponential inflation,
the Hubble radius $H^{-1}$ is constant in physical spatial coordinates
(the horizontal axis), whereas it increases linearly in time
after $t_R$. The physical length corresponding to a fixed
comoving length scale labelled by its wavenumber $k$ increases
exponentially during inflation but increases less fast than
the Hubble radius (namely as $t^{1/2}$) after inflation.}
   \label{fig:1}
   \end{figure}

A further problem is the {\it singularity problem}. It was known for a long
time that standard Big Bang cosmology cannot be the complete story of
the early universe because of
the initial singularity, a singularity which is unavoidable when basing
cosmology on Einstein's field equations in the presence of a matter
source obeying the weak energy conditions (see e.g. \cite{HE} for
a textbook discussion). Recently, the singularity theorems have been
generalized to apply to Einstein gravity coupled to scalar field
matter, i.e. to scalar field-driven inflationary cosmology \cite{Borde}.
It is shown that in this context, a past singularity at some point
in space is unavoidable. Thus we know, from the outset, that scalar
field-driven inflation cannot be the ultimate theory of the very
early universe.

The Achilles heel of scalar field-driven inflationary cosmology is,
however, the {\it cosmological constant problem}. We know from
observations that the large quantum vacuum energy of field theories
does not gravitate today. However, to obtain a period of inflation
one is using the part of the energy-momentum tensor of the scalar field
which looks like the vacuum energy. In the absence of a convincing
solution of the cosmological constant problem it is unclear whether
scalar field-driven inflation is robust, i.e. whether the
mechanism which renders the quantum vacuum energy gravitationally
inert today will not also prevent the vacuum energy from
gravitating during the period of slow-rolling of the inflaton 
field. Note that the
approach to addressing the cosmological constant problem making use
of the gravitational back-reaction of long range fluctuations
(see \cite{RHBrev4} for a summary of this approach) does not prevent
a long period of inflation in the early universe.

Finally, a key challenge for inflationary cosmology is to find a
well-motivated candidate for the scalar field which drives inflation,
the inflaton. Ever since the failure of the model of {\it old inflation}
\cite{Guth,Sato}, it is clear that physics beyond the Standard Model
of particle physics must be invoked. 

It is likely that string theory will provide ideas which allow us
to successfully address some of the above-mentioned problems of
the current versions of inflationary cosmology. Foremost, since one
of its goals is to resolve space-time singularities,
string theory has a good chance of providing a nonsingular cosmology (and {\it
string gas cosmology}, the scenario explored below, indeed has the
potential of resolving the cosmological singularity). Since string
theory should describe physics on all scales, it should be possible
to compute the spectrum of cosmological
perturbations in a consistent way within string theory, thus opening the {\it
trans-Planckian window of opportunity}. Finally, string theory contains
many light scalar fields, good candidates to be the inflaton.

Immediate problems which arise when trying to connect string theory
with cosmology are the {\it dimensionality} and {\it moduli problems}. 
Critical superstring theory
is perturbatively consistent only in ten space-time dimensions, but we
only see three large spatial dimensions. The original approach to 
addressing this problem was to assume that the six extra dimensions are
compactified on a very small space which cannot be probed with our
available energies. However, from the point of view of cosmology,
it is quite unsatisfactory not to be able to understand why it is
precisely three dimensions which are not compactified and why the compact
dimensions are stable. Brane world cosmology \cite{branereview} provides
another approach to this problem: it assumes that we live on a
three-dimensional brane embedded in a large nine-dimensional space.
Once again, a cosmologically satisfactory theory should explain
why it is likely that we will end up exactly on a three-dimensional
brane (for some interesting work addressing this issue see
\cite{Mahbub,Mairi,Lisa}). In the context of the first approach, one
must then understand why the volume and shape of the extra dimensions
are stabilized. This is the main aspect of the moduli problem. Any
time dependence of the size and shape of the extra dimensions would likely
lead to effects on our four space-time dimensional physics such as
time-dependent coupling constants or fifth forces which are not
observed. In the context of the second approach, a problem
which replaces the moduli problem is the question of why gravity
is localized precisely on our brane.

Finding a natural solution to the dimensionality and moduli problems are thus 
key challenges for superstring cosmology. These challenges have
various aspects. First, there must be a mechanism which singles out
three dimensions as the number of spatial dimensions we live in.
Second, the moduli fields which describe the volume and the shape of
the unobserved dimensions must be stabilized. As
mentioned above, solving the {\it singularity problem} is another of
the main challenges. These are the three problems which {\it string
gas cosmology} \cite{BV,TV,ABE} explicitly addresses at the present
level of development.

In order to make successful connection with late time cosmology,
any approach to string cosmology must also solve the 
{\it flatness problem}, namely make sure that the three large
spatial dimensions obtain a sufficiently high entropy (size) to
explain the current universe. Finally, it must provide a
mechanism to produce a nearly scale-invariant spectrum of
nearly adiabatic cosmological perturbations.

Since superstring theory leads to many light scalar fields, it is
 possible that superstring cosmology will provide a
convincing realization of inflation (see e.g. \cite{stringinflation}
for reviews of recent work attempting to obtain inflation in the
context of string theory). However, it is also possible that
superstring cosmology will provide an alternative to cosmological
inflation, maybe along the lines of the Pre-Big-Bang \cite{PBB}
or Ekpyrotic \cite{KOST} scenarios. The greatest challenge for
these alternatives is to solve the flatness problem (see e.g.
\cite{Pyro}). 

{\it String gas cosmology} is one approach to combining string
theory and cosmology. Its basics will be reviewed in the
following section. Recently, there has been a lot of progress
on the issue of moduli stabilization in string gas cosmology.
This is the main topic of this lecture and will be the focus
of the third section.

\section{Basics of String Gas Cosmology}

In the absence of a non-perturbative formulation of string theory,
the approach to string cosmology which we have suggested,
{\it string gas cosmology} \cite{BV,TV,ABE} (see also \cite{Perlt}
for some original work, \cite{Review2} for a short review, and
\cite{Review4} for a more comprehensive survey),
is to focus on symmetries and degrees of freedom which are new to
string theory (compared to point particle theories) and which will
be part of any non-perturbative string theory, and to use
them to develop a new cosmology. The symmetry we make use of is
{\it T-duality}, and the new degrees of freedom are 
{\it string winding modes}.

We take all spatial directions to be toroidal, with
$R$ denoting the radius of the torus. Strings have three types
of states: {\it momentum modes} which represent the center
of mass motion of the string, {\it oscillatory modes} which
represent the fluctuations of the strings, and {\it winding
modes} counting the number of times a string wraps the torus.
Both oscillatory and winding states are special to strings as
opposed to point particles. 

The energy of an oscillatory mode is independent of $R$, momentum
mode energies are quantized in units of $1/R$, i.e.
$E_m \, = \, m {1 \over R} \, ,$
and winding mode energies are quantized in units of $R$, i.e.
$E_n \, = \, n R \, ,$
where both $m$ and $n$ are integers.
The T-duality symmetry is a symmetry of the spectrum of string
states under the change
\be \label{Tdual}
R \, \rightarrow \, 1/R
\ee
in the radius of the torus (in units of the string length $l_s$).
Under such a change, the energy spectrum of string states is
invariant if winding
and momentum quantum numbers are interchanged
$
(n, m) \, \rightarrow \, (m, n) \, .
$
The string vertex operators are consistent with this symmetry, and
thus T-duality is a symmetry of perturbative string theory. Postulating
that T-duality extends to non-perturbative string theory leads
\cite{Pol} to the need of adding D-branes to the list of fundamental
objects in string theory. With this addition, T-duality is expected
to be a symmetry of non-perturbative string theory.
Specifically, T-duality will take a spectrum of stable Type IIA branes
and map it into a corresponding spectrum of stable Type IIB branes
with identical masses \cite{Boehm}.

We choose
the background to be dilaton gravity. It is crucial to include the
dilaton in the Lagrangian, firstly since
the dilaton arises in string perturbation theory at the same level
as the graviton (when calculating to leading order in the
string coupling and in $\alpha'$), 
and secondly because it is only the action of
dilaton gravity (rather than the action of Einstein gravity)
which is consistent with the T-duality symmetry. Given this
background, we consider an ideal gas of matter made up of all
fundamental states of string theory, in particular including
string winding modes.
 
Any physical theory requires initial conditions. We assume that
the universe starts out small and hot. For simplicity, we take
space to be toroidal, with radii in all spatial directions given by
the string scale. We assume that the initial energy density 
is very high, with an effective temperature which is close
to the Hagedorn temperature, the maximal temperature
of perturbative string theory. 

Based on the T-duality
symmetry, it was argued \cite{BV} that the cosmology resulting from SGC
will be non-singular. For example, as the background radius $R$
varies, the physical temperature $T$ will obey the symmetry
$ T(R) \, = \, T(1/R) \, $
and thus remain non-singular even if $R$ decreases to zero.
Similarly, the length $L$ measured by a physical observer will
be consistent with the symmetry (\ref{Tdual}), hence realizing
the idea of a minimal physical length \cite{BV}.

Next, it was argued \cite{BV} that in order for spatial sections to become
large, the winding modes need to decay. This decay, at least
on a background with stable one cycles such as a torus, is only
possible if two winding modes meet and annihilate. Since string
world sheets have measure zero probability for intersecting in more
than four space-time dimensions, winding modes can annihilate only
in three spatial dimensions. Numerical confirmation for this
scenario was provided in \cite{Sakell} (see also \cite{Cleaver}, but
see also the recent
caveats to this conclusion based on the work of \cite{Kabat3}). 
Thus, only three spatial dimensions
can become large, hence explaining the observed dimensionality of
space-time. As was shown later \cite{ABE}, adding branes to
the system does not change these conclusions since at later
times the strings dominate the cosmological dynamics.
Note that in the three dimensions which are becoming large there
is a natural mechanism of isotropization as long as some winding
modes persist \cite{Watson1}.

The equations of SGC are based on coupling an ideal gas of all
string and brane modes, described by an energy density $\rho$
and pressures $p_i$ in the i'th spatial direction, 
to the background space-time of
dilaton gravity. They follow from varying the action
\be \label{dilgrav}
S \, = \, {1 \over {2 \kappa^2}} \int d^{10}x \sqrt{-g} e^{-2 \phi}
\bigl[{\hat R} + 4 \partial^{\mu} \phi \partial_{\mu} \phi \bigr] + S_m \, ,
\ee
where $g$ is the determinant of the metric, ${\hat R}$ is the Ricci scalar,
$\phi$ is the dilaton, 
$\kappa$ is the reduced gravitational constant in ten dimensions,
and $S_m$ denotes the matter action. The
metric appearing in the above action is the metric in the
string frame. In the case of a homogeneous
and isotropic background given by
\be
ds^2 \, = \, dt^2 - a(t)^2 d{\bf x}^2 \, ,
\ee
the three resulting equations (the
generalization of the two Friedmann equations plus the equation
for the dilaton) in the string frame are
\cite{TV} (see also \cite{Ven})
\bea
-d {\dot \lambda}^2 + {\dot \varphi}^2 \, &=& \, e^{\varphi} E 
\label{E1} \\
{\ddot \lambda} - {\dot \varphi} {\dot \lambda} \, &=& \,
{1 \over 2} e^{\varphi} P \label{E2} \\
{\ddot \varphi} - d {\dot \lambda}^2 \, &=& \, {1 \over 2} e^{\varphi} E \, ,
\label{E3}
\eea
where $E$ and $P$ denote the total energy and pressure, respectively,
$d$ is the number of spatial dimensions, and we have introduced the
logarithm of the scale factor 
\be
\lambda(t) \, = \, {\rm log} (a(t))
\ee
and the rescaled dilaton
\be
\varphi \, = \, 2 \phi - d \lambda \, .
\ee

For the moment, let us focus on the effects of winding and momentum modes
individually on the evolution of a homogeneous space-time of $d=9$ spatial
dimensions. In accordance with the ideal gas approximation, we assume
that each spatial dimension has the same number of momentum and winding
modes. The contribution of the momentum modes to the pressure is positive,
with an equation of state 
\be
P \, = \, {1 \over d} E \, .
\ee
In contrast, winding modes lead to tension, i.e. negative pressure. The
equation of state of winding modes is
\be
P \, = \, - {1 \over d} E \, .
\ee
Since the energy of winding modes scales as $R$, whereas that of the momentum
modes is proportional to $1/R$, we see immediately that, as long as the number
of momentum and winding modes does not change, the momentum modes dominate for
small values of $R$, whereas the winding modes dominate for large values. If
the numbers of winding and momentum modes are equal (i.e. $n = m$), then the
critical radius, the radius where the pressure changes sign, is the string
scale, in this context called the {\it self-dual} radius.

We now see immediately from (\ref{E2}) 
that a gas of strings containing both stable winding and
momentum modes will lead to the stabilization of the
radius of the torus: windings prevent expansion, momenta
prevent the contraction. The right hand side of the equation
can be interpreted as resulting from a confining potential for
the scale factor. Note that this behavior is a consequence
of having used dilaton gravity rather than Einstein gravity
as the background. The dilaton is evolving at the time when
the radius of the torus is at the minimum of its potential.

The above background equations thus demonstrate that, in order
for any spatial dimensions to be able to grow large, the
winding modes circling this dimension must be able to
annihilate. In the case of three spatial dimensions, the
interaction of string winding modes can be described in
analogy with the interaction of cosmic strings. Two winding strings
with opposite orientations
can intersect, producing closed loops with vanishing winding
as a final state. The equations which describe the energy
transfer between winding and non-winding strings are given
in analogy to the case of cosmic strings 
(see e.g. \cite{VilShell,HK,RHBrev5} for reviews). First, we
split the energy density in strings into the density in winding
strings
\be
\rho_w(t) \, = \, \nu(t) \mu t^{-2} \, ,
\ee
where $\mu$ is the string mass per unit length, and $\nu(t)$ is
the number of strings per Hubble volume, and into the density in
string loops
\be
\rho_l(t) \, = \, g(t) e^{-3 (\lambda(t) - \lambda(t_0))} \, ,
\ee
where $g(t)$ denotes the comoving number density of loops, normalized
at a reference time $t_0$. In terms of these variables, the
equations describing the loop production from the interaction of
two winding strings are \cite{BEK}
\bea
{{d \nu} \over {dt}} \, &=& \, 
2 \nu \bigl( t^{-1} - H \bigr) - c' \nu^2 t^{-1} \label{E4} \\
{{dg} \over {dt}} \, &=& \, c' \mu t^{-3} \nu^2 
e^{3 \bigl( \lambda(t) - \lambda(t_0) \bigr)} \label{E5}
\eea
where $c'$ is a constant, which is of order unity for cosmic strings
but which depends on the dilaton in the case of fundamental strings
\cite{Kabat3}.

The system of equations (\ref{E1}, \ref{E2}, \ref{E3}, \ref{E4}, \ref{E5})
was studied in \cite{BEK} (see also \cite{Campos03}). It was
verified that the presence of a large initial density of string winding
modes causes any initial expansion of $a(t)$ to come to a halt. Thereafter,
$a(t)$ will decrease. The resulting increase in the density of winding
strings will lead to rapid loop production, and the number of winding
strings will decrease, then allowing $a(t)$ to start expanding again.
In \cite{BEK}, this initial evolution of $a(t)$ was called ``loitering''.
In \cite{BEK}, the analysis was performed using a constant value of $c'$.
Taking into account the dilaton dependence of $c'$, one finds
\cite{Kabat3} that the annihilation mechanism and resulting liberation
of the three large dimensions only works if the initial value of the
dilaton is sufficiently large.

\section{Moduli Stabilization in String Gas Cosmology}

In the following, we shall assume that either the mechanism of \cite{BV}
for setting in motion the preferential expansion of exactly three
spatial dimensions works, or, alternatively, that three dimensions
are distinguished from the beginning as being large. In either case,
we must address the moduli stabilization problem. We will discuss
recent progress on this issue in three steps. First, we study
the stabilization of the radion degrees of freedom (the radii of
the extra dimensions) in the string frame. Next, we consider the
analogous problem in the Einstein frame, and, in a third step, we
will consider the stabilization of the shape moduli.

\subsection{Radion Stabilization in the String Frame}

Let us first consider radion stabilization in the string frame
\cite{Watson2}. For this purpose, we must generalize our ansatz
for the metric to an anisotropic one
\be
ds^2 \, = \, dt^2 - e^{2 \lambda} d{\bf x}^2 - e^{2 \nu} d{\bf y}^2 \, ,
\ee
where ${\bf x}$ are the coordinates of the three large dimensions and
${\bf y}$ the coordinates of the internal dimensions.

The variational equations of motion which follow from the dilaton
gravity action (\ref{dilgrav}) for the above anisotropic metric are
\bea
- 3 {\ddot \lambda} - 3 {\dot \lambda}^2 - 6 {\ddot \nu} - 6 {\dot \nu}^2
+ 2 {\ddot \phi} \, &=& \, {1 \over 2} e^{2 \phi} \rho \\
{\ddot \lambda} + 3 {\dot \lambda}^2 + 6 {\dot \lambda} {\dot \nu}
- 2 {\dot \lambda} {\dot \phi} \, &=& \, {1 \over 2} e^{2 \phi} p_{\lambda} \\
{\ddot \nu} + 6 {\dot \nu}^2 + 3 {\dot \lambda} {\dot \nu}
- 2 {\dot \nu} {\dot \phi} \, &=& \, {1 \over 2} e^{2 \phi} p_{\nu} \\
- 4 {\ddot \phi} + 4{\dot \phi}^2 - 12 {\dot \lambda}{\dot \phi}
- 24 {\dot \nu} {\dot \lambda} + 3 {\ddot \lambda} + 6 {\dot \lambda}^2
+ 6 {\ddot \nu} + 21 {\dot \nu}^2 + 18 {\dot \lambda} {\dot \nu} \, &=& 
\, 0 \, .
\eea
where $\rho$ is the energy density and $p_{\lambda}$ and $p_{\nu}$
are the pressure densities in the non-compact and compact directions, 
respectively.

Let us now consider a superposition of several string gases, one with
momentum number $M_3$ about the three large dimensions, one with
momentum number $M_6$ about the six internal dimensions, and a further
one with winding number $N_6$ about the
internal dimensions. Note that there are no winding modes about
the large dimensions ($N_3 = 0$), either because they have already annihilated
by the mechanism discussed in the previous section, or they were never
present in the initial conditions. In this case, the energy $E$ and
the total pressures $P_{\lambda}$ and $P_{\nu}$ are given by
\bea
E \, &=& \, 
\mu \bigl[ 3  M_3 e^{- \lambda} + 6 M_6 e^{- \nu} + 6 N_6 e^{\nu} \bigr] \\
P_{\lambda} \, &=& \, \mu M_3 e^{- \lambda} \\ 
P_{\nu} \, &=& \, \mu \bigl[ - n_6 e^{\nu} + M_6 e^{- \nu} \bigr] \, ,
\label{EA3}
\eea
where $\mu$ is the string mass per unit length. In the following
subsection we will consider a more realistic string gas, a gas
made up of strings which have momentum, winding and oscillatory
quantum numbers together. The states considered here are massive,
and would not be expected to dominate the thermodynamical partition
function if there are states which are massless. However, for the
purpose of studying radion stabilization in the string frame, the
use of the above naive string gas is sufficient.

We are interested in the symmetric case $M_6 = N_6$
In this case, it follows from (\ref{EA3}) that the equation of motion for
$\nu$ is a damped oscillator equation, with the minimum of the effective
potential corresponding to the self-dual radius. The damping is due
to the expansion of the three large dimensions (the expansion
of the three large dimensions is driven by the pressure
from the momentum modes $N_3$). Thus, we see that the
naive intuition that the competition of winding and momentum modes about
the compact directions stabilizes the radion degrees of freedom at the
self-dual radius generalizes to this anisotropic setting.

\subsection{Radion Stabilization in the Einstein Frame using Massless Modes}

In order to make contact with late time cosmology, it is important to
consider the issue of radion stabilization when the dilaton is frozen, 
or, more generally, in the Einstein frame. As was discussed in 
\cite{Subodh1,Subodh2} (see also earlier comments in \cite{Watson2}),
the existence of string modes which are massless at the self-dual radius
is crucial in obtaining radion stabilization in the Einstein frame
(for more general studies of the importance of
massless modes in string cosmologys see \cite{Watson3,Eva}). Such
massless modes do not exist in all known string theories. They
exist in the Heterotic theory, but not in Type II theories \cite{Pol}.

Let us consider the equations of motion which arise from coupling
the Einstein action (as opposed to the dilaton gravity action)
to a string gas. In the anisotropic setting when the metric
is taken to be
\be
ds^2 \, = \, dt^2 - a(t)^2 d{\bf x}^2 - 
\sum_{\alpha = 1}^6 b_{\alpha}(t)^2 dy_{\alpha}^2 \, ,
\ee
where the $y_{\alpha}$ are the internal coordinates, the
equation of motion for $b_{\alpha}$ becomes
\be \label{extra}
{\ddot b_{\alpha}} + 
\bigl( 3 H + \sum_{\beta = 1, \beta \neq \alpha}^6 {{{\dot b_{\beta}}} \over {b_{\beta}}} \bigr) {\dot b_{\alpha}} \, = \, 
\sum_{n, m} 8 \pi G {{\mu_{m,n}} \over {\sqrt{g} \epsilon_{m,n}}}{\cal S} \,
\ee
where $\mu_{m,n}$ is the number density of $(m,n)$ strings, $\epsilon_{m,n}$
is the energy of an individual $(m,n)$ string, and $g$ is the determinant of
the metric. The source term ${\cal S}$ depends on the quantum numbers of the
string gas, and the sum runs over all momentum numbers and winding
number vectors $m$ and $n$, respectively (note that $n$ and $m$ are
six-vectors, one component for each internal dimension). If the number
of right-moving oscillator modes is given by $N$, then the source term
for fixed $m$ and $n$ is
\be \label{source}
{\cal S} \, = \, \sum_{\alpha} \bigl( {{m_{\alpha}} \over {b_{\alpha}}} \bigr)^2
- \sum_{\alpha} n_{\alpha}^2 b_{\alpha}^2 
+ {2 \over {D - 1}} \bigl[ (n,n) + (n, m) + 2(N - 1) \bigr] \, .
\ee
To obtain this equation, we have made use of the mass spectrum of
string states and of the level matching conditions. In the case of
the bosonic superstring, the mass spectrum for fixed $m, n, N$ and 
${\tilde N}$,
where ${\tilde N}$ is the number of left-moving oscillator states,
on a six-dimensional torus whose radii are given by $b_{\alpha}$ is
\be
m^2 \, = \, {{m_{\alpha}} \over {b_{\alpha}}} \bigr)^2
- \sum_{\alpha} n_{\alpha}^2 b_{\alpha}^2 + 2 (N + {\tilde N} - 2) \, ,
\ee
and the level matching condition reads
\be
{\tilde N} \, = \, (n,m) + N \, ,
\ee
where $(n,m)$ indicates the scalar product of $n$ and $m$ in the 
trivial trivial metric.

There are modes which are massless at the self-dual radius $b_{\alpha} = 1$.
One such mode is the graviton with $n = m = 0$ and $N = 1$. The modes of
interest to us are modes which contain winding and momentum, namely 
\begin{itemize}
\item{} $N = 1$, $(m,m) = 1$, $(m, n) = -1$ and $(n,n) = 1$;
\item{} $N = 0$, $(m,m) = 1$, $(m, n) = 1$ and $(n,n) =  1$;
\item{} $N = 0$  $(m,m) = 2$, $(m, n) = 0$ and $(n,n) =  2$.
\end{itemize}
Note that some of these modes survive in the Heterotic string theory, but
do not survive the GSO \cite{Pol} truncation in Type II string
theories.

In string theories which admit massless states (i.e. states
which are massless at the self-dual radius), these states
will dominate the initial partition function. The background
dynamics will then also be dominated by these states. To understand
the effect of these strings, consider the equation of motion (\ref{extra})
with the source term (\ref{source}). The first two terms in the
source term correspond to an effective potential with a stable
minimum at the self-dual radius. However, if the third term in the
source does not vanish at the self-dual radius, it will lead to
a positive potential which causes the radion to increase. Thus,
a condition for the stabilization of $b_{\alpha}$ at the self-dual
radius is that the third term in (\ref{source}) vanishes at the
self-dual radius. This is the case if and only if the string state
is a massless mode.

The massless modes have other nice features which are explored in
detail in \cite{Subodh2}. They act as radiation from the
point of view of our three large dimensions and hence do not
lead to a over-abundance problem. As our three spatial dimensions
grow, the potential which confines the radion becomes shallower.
However, rather surprisingly, it turns out the the potential
remains steep enough to avoid fifth force constraints. 

Key to the success in simultaneously avoiding the moduli overclosure
problem and evading fifth force constraints is the fact that
the stabilization mechanism is an intrinsically stringy one,
as opposed to an effective field theory mechanism. In the case
of effective field theory, both the confining force and the overdensity
in the moduli field scale as $V(\varphi)$, where $V(\varphi)$ is the
potential energy density of the field $\varphi$. In contrast, in
the case of stabilization by means of massless string modes, the
energy density in the string modes (from the point of view of
our three large dimensions) scales as $p_3$, whereas the confining
force scales as $p^{-1}$, where $p_3$ is the momentum in the three
large dimensions. Thus, for small values of $p$, one simultaneously
gets large confining force (thus satisfying the fifth force constraints) 
and small energy density \cite{Subodh2,Subodh3}.

\subsection{Shape Moduli Stabilization}

In the presence of massless string states, the shape moduli also
can be stabilized, at least in the simple toroidal backgrounds
considered so far \cite{Edna} (see also \cite{Sugumi}, a paper
which appeared after the Kyoto workshop). To study this issue, we consider
a metric of the form
\be
ds^2 \, = \, dt^2 - d{\bf x}^2 - G_{mn}dy^mdy^n \, ,
\ee
where the metric of the internal space (here for simplicity
considered to be a two-dimensional torus) contains a shape
modulus, the angle between the two cycles of the torus:
\be
G_{11} \, = \, G_{22} \, = \, 1
\ee
and
\be
G_{12} \, = \, G_{21} \, = \, {\rm sin}\theta \, ,
\ee
where $\theta = 0$ corresponds to a rectangular torus. The ratio
between the two toroidal radii is a second shape modulus. However,
from the discussion of the previous subsection we already know
that each radion individually is stabilized at the self-dual
radius. Thus, the shape modulus corresponding to the ratio of
the toroidal radii is fixed, and the angle is the only shape modulus which is
yet to be considered.

Combining the $00$ and the $12$ Einstein equations, we obtain
a harmonic oscillator equation for $\theta$ with $\theta = 0$
as the stable fixed point.
\be
{\ddot \theta} + 8K^{-1/2} e^{-2 \phi} \theta \, = \, 0 \, ,
\ee
where $K$ is a constant whose value depends on the quantum numbers
of the string gas. In the case of an expanding three-dimensional
space we would have obtained an additional damping term in the
above equation of motion.

We thus conclude that the shape modulus is dynamically stabilized at a value
which maximizes the area to circumference ratio. 

\subsection{Discussion}

In this section we have discussed the dynamical stabilization of
volume and shape moduli in string gas cosmology. We have seen that
simple but stringy mechanisms (not visible from an effective field
theory point of view) yield a robust stabilization mechanism for
both volume and shape moduli. The mechanism relies in a crucial
way on the existence of massless string modes involving both momentum
and winding. Neither fluxes nor non-perturbative effects are
required to stabilize these moduli fields, rendering the
stabilization mechanism much easier to analyze than the
stabilization mechanisms invoked in the case of effective field
theory flux compactification scenarios \cite{GKP}.
The stabilization mechanism is more stringy than the mechanisms
used in the work of \cite{GKP} and followup papers.

In string gas cosmology models based on a Heterotic string theory
(which has the crucial massless modes), the dilaton is not
yet stabilized in the analyses performed to date (see e.g.
\cite{BattWat,Aaron,Tirtho,Damien} for papers discussing
some of the challenges of dilaton stabilization). It appears
that new ingredients are required. In \cite{Subodh3}, ideas
for dilaton stabilization using fundamental D-branes were put
forward. However, D-branes do not exist in Heterotic theories.
In models where the dilaton can be simply stabilized using
string/brane gas cosmology ideas, the radion cannot. More work
is thus required to convincingly fix all moduli in string gas
cosmology.

The results on moduli stabilization in string gas cosmology have
been reached by use of simple stringy ingredients. This is to
be contrasted with the more involved concepts which must
be used in order to stabilize the moduli in the context of
low energy effective supergravity theories \cite{GKP} (see
e.g. \cite{Eva2} for a review with a comprehensive list of
references). On the other hand, the techniques have only
been demonstrated for simple toroidal backgrounds (for 
extensions to certain orbifolds see \cite{Easther1}).

\section{Challenges for String Gas Cosmology}

There are significant outstanding challenges for string gas
cosmology. First of all, there are basic issues of consistency
to be addressed. At the self-dual radius, higher order
corrections to the background equations will become important.
It should be checked whether the main conclusions reached to
date in string gas cosmology are robust towards these corrections.
At first sight, the answer should be ``yes'', since the main
conclusions depend on new degrees of freedom and on new
symmetries which are valid beyond perturbative string theory.

A challenge already mentioned in the previous section is
to complete the program of stabilizing all moduli fields.
In the case of models based on Heterotic string theory, the
dilaton remains to be fixed. In models based on Type II
string theory, the volume modulus is the key modulus which
remains to be stabilized.

An outstanding problem for string gas cosmology is to
establish a successful connection with late time cosmology.
One avenue would be to have a period of inflation of the
three large dimensions (for ideas on how to obtain this
see \cite{Moshe,Easson2,Tirtho2}). However, the danger in
such a scenario is that the number density of strings winding
the extra dimensions is diluted to an extent inconsistent with
moduli stabilization. Thus, starting from initial conditions where
all spatial dimensions are small, the challenge is to obtain
inflation consistent with moduli stabilization. Obviously, it
is possible that our three spatial dimensions were always
sufficiently large, even when the density of strings winding the
extra dimensions was large. Novel ideas for late-time
cosmology in this context were recently put forwards in
\cite{Gubser,Tirtho3}.

It is possible that string gas cosmology will connect with
the late time universe without a period of conventional
inflation. In this case, a new mechanism for the origin of
cosmological fluctuations would be required. A key challenge
in this context is to find a solution of the entropy problem,
namely to explain why our three spatial dimensions are so large
\cite{Pyro}.

In conclusion, we have shown that string gas cosmology provides
the potential for yielding a non-singular cosmology, a cosmology
which explains why only three spatial dimensions are very large,
and why the moduli describing the volume and shape of the extra
dimensions are stabilized. Much more work remains to be done
to put string gas cosmology on a more consistent mathematical
basis (see \cite{list} for some recent papers on string gas
cosmology which were not explicitly discussed in this brief
review).

\centerline{\bf Acknowledgements}

I would like to thank the organizers of the YKIS Conference
(and in particular Profs. M. Sasaki and H. Kodama) for inviting me to speak
and for their wonderful hospitality during my visit in Kyoto. I
wish to thank my graduate students Thorsten Battefeld, Subodh
Patil and Scott Watson for their insights. Most of the results
reported in this lecture are based on their work. This research is
supported by an NSERC Discovery Grant and by the Canada Research
Chairs program.

\end{document}